\documentclass[twocolumn,showpacs,preprintnumbers,amsmath,amssymb]{revtex4}

\usepackage{graphicx}
\usepackage{dcolumn}
\usepackage{bm}

\begin{document}

\preprint{APS/123-QED}

\title{Gutzwiller study of spin-1 bosons in an optical lattice 
under a magnetic field}

\author{Takashi Kimura,${}^1$ Shunji Tsuchiya,${}^2$ 
Makoto Yamashita,${}^3$ and Susumu Kurihara${}^4$}
 \affiliation{${}^1$Department of Information Science, Kanagawa University, 
2946 Tsuchiya, Hiratsuka 259-1293, Japan\\ 
${}^2$Institute of Industrial Science, University of Tokyo, 
4-6-1 Komaba, Meguro, Tokyo 153-8904, Japan\\
${}^3$NTT Basic Research Laboratories, 3-1 Morinosato-Wakamiya, 
Atsugi 243-0198, Japan\\
${}^4$Department of Physics, Waseda University, 3-4-1 Ohkubo, Shinjuku, Tokyo 169-8555, Japan
}%

\date{\today}

\begin{abstract}
We study spin-1 bosons in an optical lattice under a magnetic field
by the Gutzwiller approximation.
Our results thus obtained 
join the discontinuous phase boundary curves obtained by perturbative studies
through a first-order transition.  On the phase boundary curve,
we also find a peculiar cusp structure 
originating from the degeneracy between different spin Mott states 
under a magnetic field. 
The magnetic field dependence of both fluctuation in the total number
of bosons and the spin magnetization 
clarifies that the superfluid phase is divided into two regions reflecting 
the coexisting first- and second-order superfluid transitions. 
\end{abstract}

\maketitle 
Superfluid (SF) and Mott insulators (MI) are very interesting 
subjects in condensed matter physics. 
Recently, the transition between the 
SF and MI of spinless Bose atoms 
has been experimentally demonstrated in an optical lattice system \cite{Greiner}
closely following the theoretical proposal 
based on a Bose-Hubbard model (BHM)
\cite{Fisher,BHM} due to Jaksch et al. \cite{Jaksch}. 
In contrast to the case of  fermions \cite{Review}, 
the Gutzwiller approximation (GA) \cite{BoseGutz} 
describes an accurate second-order SF-MI transition of spinless bosons, 
which is supported by quantum Monte Carlo simulations \cite{QMC}.

Trapped spinor bosons have also been investigated
both theoretically \cite{spintheory} and experimentally \cite{spinorexp}.
In an optical lattice, MI\cite{DemlerZhou,nematic1} 
and SF\cite{DemlerZhou,Cheng} phases of spinor bosons 
have been studied. MI phases under a magnetic field 
have also been studied recently \cite{magMI}. 
The SF-MI transition in a spin-1 BHM 
has also been studied 
by a perturbative mean-field approximation (PMFA)
\cite{Oosten},  
which treats the hopping process between adjacent sites as a perturbation  
\cite{Tsuchiya}, by the GA \cite{Kimura}, by a conventional 
mean-field approximation (CMFA) \cite{Krutitsky}, and by 
density-matrix renormalization group in one dimension \cite{Rizzi}. 
The GA and CMFA results show a possible first-order transition (FOT)
at a part of the phase boundary curve, 
where the critical value of hopping matrix element 
just on the phase boundary 
is smaller than that obtained by the PMFA,  
which describes a second-order transition (SOT). 
On the other hand, the results obtained by the PMFA
can also be obtained by the GA, which assumes 
a partial set of the full wave functions \cite{Kimura}. 
Hence, generally speaking, the results obtained by  
the GA improve those obtained by the PMFA. 

The PMFA approach has also been applied to 
the SF-MI transition 
in the spin-1 BHM under a magnetic field \cite{Uesugi,Svidzinsky}.
However, the phase boundary curve obtained by the PMFA 
in the limit of a weak magnetic field 
($B\rightarrow 0$) \cite{Uesugi,Svidzinsky}
does not agree with that obtained 
by initially assuming zero magnetic field 
($B=0$) \cite{Tsuchiya}
when the MI phase has an odd number of bosons. 
Moreover, the phase boundary obtained by the PMFA 
\cite{Uesugi,Svidzinsky} is a discontinuous 
function of magnetic field. 
These features might be attributed to the PMFA itself. 

In this paper, we study the spin-1 BHM by the GA. 
In contrast to the PMFA, the GA shows that 
the SF-MI phase boundary is a continuous 
function of magnetic field even around a zero magnetic field. 
In the phase diagram, we also find a special point 
where a degeneracy of MI states with different spins plays an important role.
We also investigate superfluid properties in terms of magnetization and 
fluctuation in the total number of bosons (FTNB) in the system. 
Both quantities, which are experimentally observable, 
have interesting magnetic field dependence 
originating from the coexisting FOT and SOT 
in the present system.

The BHM of spin-1 bosons \cite{DemlerZhou,Tsuchiya,Kimura,nematic1,Uesugi,Svidzinsky} 
is given by $H=H_{\mathrm hop}+H_{\mathrm int}+H_{\rm mag}$ in standard notation \cite{Tsuchiya} 
\begin{eqnarray}
H_{\mathrm hop}&=&-t\sum_{\langle i,j \rangle}\sum_{\alpha=0,\pm1}({\hat a}_{i \alpha}^\dagger {\hat a}_{j \alpha}^{}
 + {\hat a}_{j \alpha}^\dagger {\hat a}_{i \alpha}^{}), \nonumber\\
H_{\mathrm int}&=&\sum_i\big[-\mu{n}_i+\frac{U_0}{2}{n}_i({n}_i-1)
+\frac{U_2}{2}({\mathbf S}_i^2-2n_i)\big],\nonumber\\
H_{\mathrm mag}&=-&g\mu_BB\sum_i{S_{zi}}\equiv -b\sum_i{S_{zi}}.
\end{eqnarray}
Here, for simplicity, we assume that a magnetic field is parallel to the $z$ axis 
and that the system is uniform. 
In this paper, we use $U_0$ as a unit of energy 
and assume an antiferromagnetic interaction 
$U_2=0.04$, which corresponds to ${}^{23}{\mathrm Na}$ atoms.  

The Gutzwiller variational wave function for a site
is defined as $\Phi=\sum_{N}g(N)|N\rangle$, 
$|N\rangle=\sum_Sf(N,S)|N,S\rangle$, and 
$|N,S\rangle=\sum_{S_z}l(N,S,S_z)|N,S,S_z\rangle$. 
Here, $|N,S,S_z\rangle$ has $N$ bosons, a total spin $S$, 
and a magnetic quantum number $S_z$, where
$S$ must be odd (even) for an odd (even) $N$ \cite{DemlerZhou}.
The variational parameters must satisfy the  normalization condition 
$\sum_{N} |g(N)|^2=\sum_S |f(N,S)|^2=\sum_{S_z}|l(N,S,S_z)|^2=1$.  
In this paper, we take the complete set from $N=0$ to $N=6$,  
which is sufficient for a numerical convergence. 
We employ a standard definition such that 
the MI phase has zero FTNB
$\langle ({N_{\mathrm tot}}-
\langle {N}_{\mathrm tot}\rangle)^2 \rangle/
\langle {N}_{\mathrm tot}\rangle$ 
(${N}_{\mathrm tot}\equiv\sum_{i}{n}_i$),  
while the SF phase has a finite one \cite{totalnumber}. 
In an SF phase close to an MI phase with $N$ bosons, 
we consider $g(M)$ for $M\neq N$ 
as an SF-order parameter because finite $g(M)$ 
results in finite FTNB. 
The SF order parameters have finite (no) jumps 
at the SF-MI phase boundary for a FOT (SOT).

{\it Superfluid-Mott-insulator transition---}
Figure 1 shows phase boundary curves 
on the $zt$-$\mu$ plane ($z$: the number of adjacent sites)
under a very weak magnetic field $b=0.001$. 
In addition to the results obtained by the PMFA and the GA, 
we also plot those obtained by the PMFA under zero magnetic field
(dot-dashed curves). 

\begin{figure}
\includegraphics[height=2.2in]{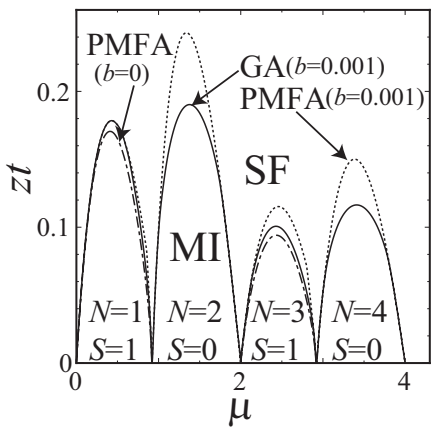}
\caption{
Phase boundary curves  
as a function of chemical potential $\mu$. 
SF and MI indicate the superfluid and the Mott-insulating phases, 
respectively. The solid and dashed  
curves are obtained for $b=0.001$ 
using the GA and the PMFA, respectively. 
The dot-dashed curves around the MI phase with $N=1,3$ 
are obtained by the PMFA under zero magnetic field
and are not visible around the MI phase with $N=2,4$ 
because they are indistinguishably close to the dashed curves. 
}
\label{fig1}
\end{figure}

We can easily see that the results obtained by the PMFA for 
$b=0$ and $b=0.001$ are clearly different from 
each other around an MI phase with an odd number of bosons. 
This difference originates from the calculation procedure. 
Namely, a degenerate PMFA
is employed to lift the degeneracy among $S_z=\pm1,0$ states 
under zero magnetic field \cite{Tsuchiya}, 
while the degeneracy has already been lifted 
under a finite magnetic field \cite{Uesugi,Svidzinsky}.
Because the PMFA neglects 
low-spin states 
even under a weak magnetic field, the PMFA 
overestimates the antiferromagnetic interaction 
energy in a possible SF phase, 
resulting in a large critical value of $zt$ for the SF-MI transition. 
The results obtained by the GA 
fall in between the two results obtained by the PMFA.     
The SF-MI transition is a FOT at a part of the phase boundary, 
where the results obtained by the GA do not completely 
agree with those obtained by the PMFA.   
In the limit of $b\rightarrow0$, 
the results obtained by 
the GA are not below those obtained by 
the PMFA initially assuming $b=0$, but completely agree with them. 
This is consistent with 
the previous result under zero magnetic field 
\cite{Kimura} such that 
a FOT only occurs for very small $U_2$ around MI phases with 
odd $N$ bosons and does not occur for any $U_2$ when $N=1$.

On the other hand, around the MI phases with an even number of bosons, 
the phase boundary curves obtained by the GA (the PMFA) for
$b=0.001$ are almost indistinguishably close to those for $b=0$,  
which are from Ref. \cite{Kimura} (Ref. \cite{Tsuchiya}). 
This shows that the singlet MI phases are robust under a weak 
magnetic field. 

Figure 2 shows the magnetic field dependence of 
the phase boundaries for $\mu=2.5$ and $\mu=1.85$.  
For $\mu=2.5$, 
the PMFA assumes an MI state with $N=3$
has a spin ($S=1$ or $S=3$) 
that discontinuously depends on 
whether $b$ is smaller or larger than 0.1 \cite{nonzero}. 
This results in a curious jump of the critical value of $zt$ for 
the SF-MI transition just for $b=0.1$. 
The solid curve obtained by the GA 
gives a continuous phase boundary by joining 
the separated dashed curves and  
exactly agrees with the dashed curves except for a finite region 
$0.0838< b<0.1$, where the transition is a FOT.  

\begin{figure}
\includegraphics[height=2.2in]{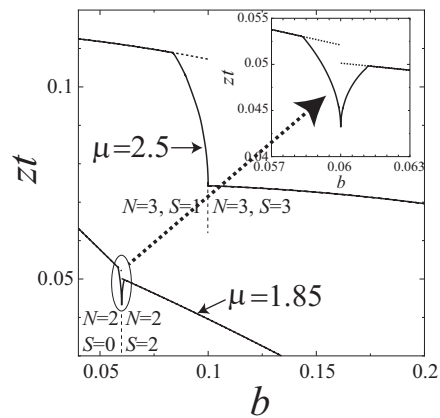}
\caption{
Phase boundary curves on a $b$-$zt$ plane for $\mu=2.5$ 
and for $\mu=1.85$.  
The dashed (solid) curves are obtained by the PMFA (GA). 
The inset is an enlargement around $b=0.06$ for $\mu=1.85$.  
}
\label{fig2}
\end{figure}

On the other hand, 
for $\mu=1.85$,  
the solid curve obtained by the GA has a sharp cusp structure
around $b=0.06$ where the MI states 
with 
$S=0$ and $S=2$ are degenerated. 
The critical value of $zt$ just for $b=0.06$ 
is different from both of the two limiting values 
from the weak or strong magnetic field region obtained by the PMFA. 
The transition is a FOT near (but not just for) $b=0.06$. 
However, the SF order parameters become smaller when 
$b$ becomes closer to $0.06$, and finally 
the transition becomes a SOT just for $b=0.06$. 
In fact, by using 
a degenerate PMFA that assumes two MI states 
such that $|N,S-2\rangle$ and $|N,S\rangle$ 
 as zeroth-order states,  
we obtain the critical value of $zt$ as  
$(zt_{\mathrm deg})^{-1}=\frac{1}{2}
\Big[a_{+} -\frac{a_-c}{2b_-}+\frac{1}{2}(1+\frac{a_{-}^2}{|b_-|})
\sqrt{\frac{c^2-4b_-b_+}{a_{-}^2-b_-}}\Big]$, where
$a_{\pm}\equiv(\alpha_{S,+}\pm\alpha_{S,-}+\alpha_{S-2,+}\pm\alpha_{S-2,-})/2$, 
$b_{\pm}\equiv[\alpha_{S,+}\pm\alpha_{S,-}
-(\alpha_{S-2,+}\pm\alpha_{S-2,-})]^2/4\pm\beta^2$, and 
$c\equiv[(\alpha_{S,+}-\alpha_{S-2,+})^2-(\alpha_{S,-}-\alpha_{S-2,-})^2]/2$. 
Here, 
$\alpha_{S,\pm}\equiv f_{N,S,S}^{N+1,S+1,S\pm1}+f_{N,S,S}^{N\mp1,S\pm1,S-1}
+f_{N,S,S}^{N-1,S\mp1,S\mp1}$ and $\beta=
\frac{{\langle N+1,S-1,S-1|{\hat a}_{-1}^\dagger|N,S,S\rangle}
\langle N,S-2,S-2|{\hat a}_1|N+1,S-1,S-1\rangle}{
E(N+1,S-1,S-1)-E(N,S,S)}+ 
\frac{{\langle N-1,S-1,S-1|{\hat a}_{1}|N,S,S\rangle}
\langle N,S-2,S-2|{\hat a}_{-1}^\dagger|N-1,S-1,S-1\rangle}{
E(N+1,S-1,S-1)-E(N,S,S)}$. 
Here, 
$E(i,j,k)$ is the energy per site of an MI state $\Phi=|i,j,k\rangle$ and   
$f_{l,m,n}^{i,j,k}\equiv \frac{
|\langle l,m,n|{\tilde a}|i,j,k\rangle|^2}{E(i,j,k)-E(l,m,n)}$, 
where ${\tilde a}$ is a creation or an annihilation 
operator which joins $|l,m,n\rangle$ and $|i,j,k\rangle$ \cite{matrix}.       
The PMFA chooses $zt_{\mathrm deg}$ 
as the critical value only when $zt_{\mathrm deg}$ 
is smaller than the two limiting values 
from the weak or strong magnetic field. 
Furthermore, another condition 
$\Big| \big[c-a_-{\rm sgn}(b_-)
\sqrt{(c^2-4b_-b_+)/(a_-^2-b_-)}\big]/(2b_-)\Big|<1$ 
must also be satisfied 
because the absolute values of the SF-order parameters must be 
non-negative. 
The latter condition is not satisfied in the case of $\mu=2.5$ in Fig. 2.
It should also be noted that 
the same critical value as $zt_{\mathrm deg}$ 
can be not only numerically but 
also analytically obtained by the GA including the states 
that emerge as zeroth-order or intermediate states
in the degenerate PMFA calculation.

{\it Superfluid properties---}
Figure 3 shows magnetization per site 
and FTNB as a function of magnetic field 
for $\mu=1.92$ and for a constant 
$zt=0.02$ \cite{parametersets}. 
The FTNB is proportional to the experimental observables, such as 
the compressibility and the inverse square of the sound velocity 
\cite{Fisher}. 
Both the magnetization and the FTNB curves 
can be clearly divided into four parts
depending on the magnetic field. 
There are always discontinuous jumps 
of the differential magnetic susceptibility 
and the derivative of the FTNB  
on the boundaries between the four parts. 

To clearly understand the magnetization and the FTNB curves, 
we also show phase boundary curves in Fig. 4,  
where the SF-MI transition is a SOT (FOT) under a magnetic field 
$b\le0.06$ or $0.0670<b$  ($0.06<b<0.0670$). 
The MI states $|N=2,S=0\rangle$ 
and $|N=2,S=2\rangle$ are degenerated at 
$b=0.06$ and the former (latter) is more favored under a 
weaker (stronger) magnetic field. 
For $zt=0.02$ in Fig. 4, 
the first region ($b<0.0462$) 
and the fourth region ($0.0612<b$) 
correspond to the MI phases 
$|N=2,S=0\rangle$ and $|N=2,S=2\rangle$, respectively,  
where the magnetization is constant and 
the FTNB is zero as shown in Fig. 3. 

The SF phase in the second region $0.0462<b<0.0596$ 
has a {\it perturbative character}: 
The SF state is continuously connected to the nearest 
MI state, and the spin property of the system is close to that of 
the MI state. 
The inset of Fig. 4 indeed shows that $|f(2,2)|^2$, 
which is the amplitude of a high spin state 
$|N=2,S=2\rangle$ 
normalized as $|f(2,0)|^2+|f(2,2)|^2=1$,   
is almost negligible 
and that the lowest spin state $|N=2,S=0\rangle$ is dominant 
in the second region. 
This perturbative character originates from 
a SOT between the SF phase in the second region 
and the MI phase in the first region. 

On the other hand, the SF in the third region has a 
{\it non-perturbative character} such that 
states with high spins 
are largely included, 
resulting in the large $\langle S_z\rangle$ as shown 
in Fig. 3. The inset of Fig. 4 shows that 
$|f(2,2)|^2$ is indeed large in the third region. 
This change of the SF character also affects the FTNB as shown in Fig. 3. 
It should also be noted the non-perturbative SF phase 
can also be characterized as a {\it coherent-state-like character} 
with a large kinetic energy as in the case of a zero magnetic field \cite{Kimura}. 
This non-perturbative character is related to 
a FOT between the SF phase in the third region
and the MI phase in the fourth region.

\begin{figure}
\includegraphics[height=2.2in]{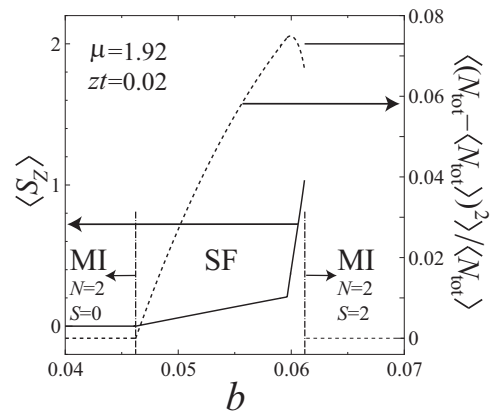}
\caption{Magnetization per site 
$\langle S_z\rangle$ (solid curves)
and fluctuation in the total number of bosons 
$\langle (N_{\mathrm tot}-\langle N_{\mathrm tot}\rangle)^2 \rangle/\langle N_{\mathrm tot}\rangle$ (dashed curves) 
as a function of magnetic field for $\mu=1.92$. 
}
\label{fig3}
\end{figure}

\begin{figure}
\includegraphics[height=2.2in]{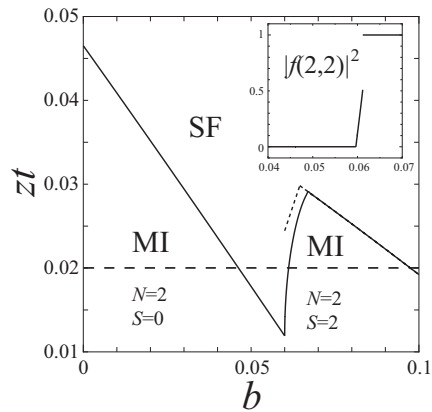}
\caption{Phase boundary curves for $\mu=1.92$.  
The solid (short dashed) curve is obtained by the GA (PMFA). 
The long dashed line, which shows the $zt=0.02$ assumed in Fig. 3, 
intersects with the solid curve 
at $b=0.0462$ and $b=0.0612$. 
The inset shows $|f(2,2)|^2$ as a function of 
magnetic field (See text for the definition). 
}
\label{fig4}
\end{figure}

\begin{figure}
\includegraphics[height=2.2in]{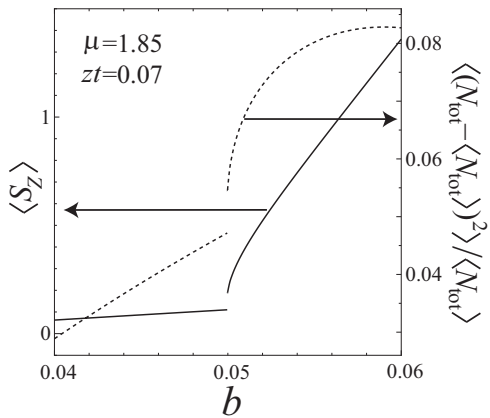}
\caption{The same plot as in Fig. 3 for $\mu=1.85$ 
and $zt=0.07$. 
}
\label{fig5}
\end{figure}

We can see the crossover between the two SF phases 
in a wide parameter region when the SOT and FOT phase boundary curves 
coexist. For instance, 
the critical value of $zt$ for the SF-MI transition 
just at the boundary between MI phases with different spins
has not to be necessary a local minimum as a function of magnetic field.  
Although not shown here, 
we indeed found a clear crossover 
between the non-perturbative and perturbative SF phases 
for $zt>0.075$ and $\mu=2.5$ around $b=0.1$ 
(See Fig. 2 for the phase boundary curve). 
Here, the critical value of $zt=0.075$ for $b=0.1$ 
is not a local minimum as a function of magnetic field. 

Let us explain some details. 
There are no gaps on the curves of 
the magnetization and the FTNB 
at the boundary between 
the perturbative and non-perturbative SF phases in Fig. 3. 
On the other hand, there are
gaps in the parameter region 
as shown in Fig. 5 for $\mu=1.85$ and $zt=0.07$, 
for which the phase boundary curves are shown in Fig 2. 
For $\mu=1.85$ and $zt=0.07$, the SF-MI transition occurs 
only once for $b\simeq0.0274$ through a SOT. 
However, a FOT also occurs for 
a slightly larger $b$ and a slightly smaller $zt$, 
resulting in a crossover between the two kinds of SF phases
for $zt=0.07$ \cite{gap}.


We finally note that the GA neglects 
inter-site correlation effects, which 
could be important when $U_2$ is smaller 
than or comparable with $zt^2/U_0$ 
and/or when the dimension of the lattice is low. 
Although $zt^2/U_0$ is somewhat smaller or much smaller 
than $U_2$ in the typical parameter sets we have assumed in this paper 
\cite{parameter}, a somewhat larger $U_2$ would more
clearly justify the GA. 
A larger $U_2$ is also favorable from the experimental point of view.
This is because both Zeeman energy and antiferromagnetic interaction
energy become comparable under a stronger magnetic field and the
interesting features clarified in this paper can be easily observed.

This work was supported by 
a Grant-in-Aid for the 21st Century COE Program of Waseda University 
(Physics of Systems with Self-Organization Composed of Multi-Elements). 



\begin{thebibliography}{18}
\bibitem{totalnumber} 
The FTNB is equal to the fluctuation of number of bosons 
in a single site as $\langle ({n}_i-\langle {n}_i\rangle)^2 \rangle/
\langle {n}_i \rangle$
within the GA.
\bibitem{nonzero} The critical value of $zt$ is generally finite 
at the phase boundary between MI states with the same 
$N$ and with different spins 
as shown in Fig. 2 because the FTNB is zero for sufficiently small $zt$.
\bibitem{parametersets} The situation is simplified in this parameter set; 
there is no FOT under a weak magnetic field: 
the degenerate PMFA is not needed in order to determine the phase boundary.
\bibitem{gap}As far as the authors know, the gaps can be finite only when 
the SOT and FOT coexist on the phase boundary around the same MI state 
as in the case for $\mu=1.85$ and $zt=0.07$.
\bibitem{parameter} 
We have typically assumed $zt<0.1$ in this paper. 
This leads to $zt^2/U_0<0.002\ll U_2$ for a three-dimensional 
cubic lattice ($z=6$).
\end{thebibliography}
\end{document}